\documentclass[11pt]{article}

\usepackage[margin=1in]{geometry}
\usepackage{parskip} 
\usepackage{etoolbox}
\apptocmd{\thebibliography}{\setlength{\itemsep}{\parskip}}{}{}
\usepackage{amsmath}
\usepackage{amssymb}
\usepackage{amsthm}
\usepackage{graphicx}
\usepackage{cite}
\usepackage{url}
\usepackage{refcount}
\usepackage{hyperref}

\newcommand{\eg}{\emph{e.g.}}
\newcommand{\ie}{\emph{i.e.}}
\newcommand{\etc}{\emph{etc.}}
\newcommand{\etal}{\emph{et~al.}}

\theoremstyle{plain}
\newtheorem{theorem}{Theorem}
\newtheorem{lemma}{Lemma}

\newtheorem*{rthm}{Theorem \ref{thm:max-es} (Maximizing $\mathrm{E}[S]$)}
\newtheorem*{rlemA}{Lemma \ref{lem:reliability-threshold} (Reliability-Threshold Relation)}

\title{Effective Interventions Against AI-Enhanced Scams}

\author{Kyle Fredrickson\\
\textit{Quarry Intelligence}}

\date{}

\begin{document}
\maketitle

\begin{abstract}
In 2025, scams were responsible for an estimated \$442 billion in direct losses globally. In the United States, reported losses increased by nearly 400\% between 2020 and 2025. Though AI in scamming is a relatively new phenomenon, its use significantly changes the economics of scams as well as the bottlenecks in scam operations. In this paper I investigate what interventions will remain effective under this new AI-driven scamming regime.

I develop a simple model of scam profits to understand how different interventions asymptotically affect scam operations. I find that three levers---reporting rate, centralization of reporting, and report accuracy---multiply in their effect on expected victims per scam channel, reducing revenue per scam channel while increasing costs. Because effects multiply, interventions affecting all three could have a significant effect on the profitability of the scam business model. My analysis suggests that even modest reporting rates against high-value scam infrastructure could have significant impacts on scam profitability.
\end{abstract}

\section{Introduction}
\label{sec:introduction}

In 2025, the Global Anti-Scam Alliance (GASA) estimated that scams were responsible for \$442 billion in direct losses \cite{2025-gasa-global-state-of-scams}. In the United States, reported losses, which are estimated to undercount real losses by ${\sim}7\times$ \cite{cfa2026_scam_economy}, have risen nearly 400\% between 2020 (\$4.2 billion lost) and 2025 (\$21 billion lost) \cite{2025-fbi-ic3-internet-crime-report}. This does not account for administrative costs, which are estimated to lie between \$5 and \$6 per reported dollar lost \cite{2025-lexisnexis-true-cost-of-fraud-north-america}. Though AI-enabled scams---those employing large language models---are a relatively new phenomenon, Chainalysis estimates they are $4.5\times$ more profitable than scams that do not use AI \cite{2026-chainalysis-fraud-report}.

AI increases scam profitability both by increasing revenue and decreasing costs \cite{2026-gressel-ai-romance-scams}. Replacing or augmenting human scammers with AI systems increases revenue by making scams more believable. The more believable a scam is, the greater the likelihood that a target who sees a \emph{scam lure}, for example, a ``wrong number'' message initiating a pig-butchering scam \cite{2025-oak-anna}, or a YouTube livestream advertising a crypto scam, will eventually convert into a victim who pays out. Scam costs can also be reduced as AI reduces the marginal cost of engaging with targets. Before AI, engaging with targets was labor intensive. In pig-butchering scams, for example, human scammers would build trust with victims for weeks or months before finally perpetrating the fraud. With AI this can be automated at negligible cost.

Scams are already a major and growing problem, and with AI we can expect the problem will accelerate. It is therefore critical that we design and implement interventions that will remain effective even after AI-enabled scams become widespread. Independently of how AI affects scams in the future, scams (1) are a business and (2) will depend on communication and financial infrastructure in order to contact victims and receive and launder their funds. I therefore focus on interventions affecting scam profitability via scam infrastructure.

\textbf{Contributions.} I develop a simple model of scam profits under worst-case assumptions about how AI will affect the economics of scamming. I underestimate scammers' total costs and consider only the cost of the communication and financial channels they use to engage with targets---the phone numbers, social media accounts, emails, and financial accounts. This is a lower bound on costs and holds regardless of exactly how AI reshapes the economics of scamming.

I use this model to estimate the effects of various interventions on scam operations. The goal is not to exactly predict the outcomes of scams, rather it is to understand how different interventions affect scam operations, profits, and costs asymptotically. I find that three levers---reporting rate, centralization of reporting, and report accuracy---lead to a multiplicative reduction in the expected victims per channel. This simultaneously leads to a multiplicative reduction in revenue per contact channel and an increase in scammers' infrastructure costs and demand for infrastructure. Interventions that even weakly affect all three of these parameters could therefore compound into a significant attack on scammers' profit margins.

Finally, I discuss the implications of this work for anti-scamming strategies and derive an upper bound on the reporting rates that would make scams unprofitable. I find that even modest reporting rates against high-cost infrastructure could have significant impacts on the profitability of scams.

\section{Background}
\label{sec:background}

Scams are distinct from other types of fraud in that scammers, under false pretenses, convince victims to willingly transfer money to them.

Scams begin with a lure sent via some communication channel, \eg, social media, SMS, email, \etc, to establish contact with their target. Scammers then engage with the target to convince them to pay. Only after this do they reveal the financial channel to be used for payment. The specifics of each step vary with each type of scam. For example, scammers may send targets SMS messages claiming they missed jury duty and there is a warrant out for their arrest, before suggesting they pay a ``fine'' \cite{2026-ftc-jury-duty-scam}. Or they may send innocuous-seeming ``wrong number'' messages in order to build a relationship with a target before introducing an investment scam \cite{2025-oak-anna}.

In the course of a scam, scammers reveal their infrastructure to many people who recognize it is being used for scams. Scammers reuse the same communication and financial channels---phone numbers, email addresses, social media accounts, money laundering accounts---across thousands of scam attempts to reduce infrastructure costs. Because scams convert at very low rates \cite{2008-kanich-spamalytics, 2024-liu-give-and-take}, thousands of people are aware that a particular channel is involved in a scam before money ever moves. In the context of YouTube livestreams advertising crypto scams, Liu \etal \cite{2024-liu-give-and-take} estimated that there were only 4 victims per 100,000 views, while each livestream averaged about 6,000 views. Even if most of those viewers did not recognize that it was a scam, there were still \emph{thousands} who knew it was.

However, the structure of scams ensures only disposable communication infrastructure is revealed while protecting the expensive financial infrastructure. Even if scammers' communication infrastructure is reported and banned, it is generally easy and cheap to replace, costing just a few dollars per account \cite{2024-beluri-sale-of-social-accounts, 2026-bulk-gmail-account-vendor}. In contrast, the financial infrastructure scammers use to move money is difficult to acquire and can cost hundreds of dollars \cite{2026-group-ib-cloud-phones, 2026-group-ib-mule-account-creation, 2025-sumsub-money-mules}. By gating their financial infrastructure behind engagement, scammers ensure their financial infrastructure is revealed primarily to their victims.

Still, the problem with scams is not that the information to stop them does not exist or that it is inaccessible. Spokoyny \etal\ \cite{2025-spokoyny-victim-as-a-service} found that scammers reveal their financial infrastructure to targets after just a few dozen messages. The problem is the parties who hold scam intelligence do not lose money to them and so have little reason to uncover or share intelligence: the incentives are misaligned.

Assuming one could correct these incentives and build an effective system to report scam infrastructure, I therefore ask: could reporting significantly affect the profitability of scams?

\section{Modeling Scam Profits}
\label{sec:modeling-scam-profits}

I model the profit of a scammer who uses some number of channels---the individual pieces of scammers' infrastructure like phone numbers, social media accounts, emails, and financial accounts---to contact victims. In this model, scammers use each channel to send \emph{lures} to victims, \eg, ``wrong number'' SMS messages that become the basis of pig-butchering scams\footnotemark[\getrefnumber{fn:pig-butchering}], or YouTube livestreams advertising crypto scams. Lures convert into victims or reports against the channel. After some number of reports, set by a platform's tolerance for false positives, the channel is banned and the lures fail.

My cost model significantly underestimates the costs of scamming in order to model a scammer's best-case scenario. I assume that with AI the cost of engaging targets will go to zero, and only consider the costs of the channels scammers use to contact victims. In reality, costs are much higher, and include fees for money laundering, payoffs to corrupt officials, labor, real estate for scam compounds, and other costs \cite{2011-levchenko-click-trajectories}.

\subsection{Scam Revenue Model}
\label{sec:scam-revenue}

The total revenue of a scam is the sum of each victim's losses, so the revenue is the average loss per victim, $\ell$, multiplied by the number of victims:
\[
\mathrm{Revenue} = \ell \cdot \overbrace{\mathrm{E}[S] \cdot m}^{\text{victims}},
\]
where $\mathrm{E}[S]$ is the expected number of victims per channel and $m$ is the number of channels.

Let $n$ be the number of lures sent per channel. Let $S$ be a random variable denoting the number of victims from $n$ lures, where $s$ is the probability of a lure converting to a victim, and let $R$ be a random variable denoting the number of reports from $n$ lures, where $r$ is the probability that a lure converts into a report against the channel. I assume for simplicity that lures convert to victims and produce reports independently, and with identical probabilities across lures. So $R \sim \mathrm{Binomial}(n, r)$. After $t$ or more reports, the channel is banned, $\Pr[S > 0 \mid R \ge t] = 0$, so for $k > 0$
\[
\Pr[S = k] = \Pr[S = k \mid R < t]\Pr[R < t].
\]

By the assumption that the probability a lure converts to a victim is independent of reporting, $\{S \mid R < t\} \sim \mathrm{Binomial}(n, s)$, and so
\[
\mathrm{E}[S] = sn \cdot \Pr[R < t].
\]

\subsection{Scam Cost Model}
\label{sec:scam-cost}

Previously, the cost of engaging victims through lures was a significant cost and bottleneck in scamming \cite{2026-heiding-ai-voice-phishing}. In the case of pig-butchering scams, a human would make initial contact with a target, then manually engage them over the course of weeks or months to build trust before finally perpetrating the fraud \cite{2025-oak-anna}. With AI this process can now be automated for negligible cost \cite{2026-gressel-ai-romance-scams}.

Regardless of how AI changes the way scammers interact with targets, what remains as a cost is the communication infrastructure used to contact targets and the financial infrastructure used to receive and launder scam proceeds. I underestimate costs and consider only the costs of the infrastructure:
\[
\mathrm{Cost} = \kappa \cdot m,
\]
where $m$ is the number of channels and $\kappa$ is the average price per channel.

\subsection{Maximizing Profits}
\label{sec:maximizing-profits}

Maximizing scam profits implies maximizing the expected victims per channel, $\mathrm{E}[S]$. The cost of acquiring a channel is a fixed overhead, and under the new AI-driven scamming regime I assume the marginal cost of sending lures and engaging victims is negligible. Thus the only force limiting how many lures a channel sends is the risk of the channel being banned. Because cost is independent of the number of lures, the number of lures that maximizes revenue also maximizes profit.

I approximate the maximum expected victims, $\mathrm{E}^\ast(s, r, t)$, attained at the optimal number of lures $n^\ast$.

\begin{theorem}[{Maximizing $\mathrm{E}[S]$}]
\label{thm:max-es}
Let $\mathrm{E}^\ast(s, r, t)$ be the maximum value of $\mathrm{E}[S \mid n]$. Then the maximum of $\mathrm{E}[S \mid n]$ is attained at $n^\ast \approx \frac{\lambda^\ast(t)}{r}$, where $\lambda^\ast(t)$ maximizes $\lambda \Pr[\mathrm{Poisson}(\lambda) < t]$, and
\[
\mathrm{E}^\ast(s, r, t) \approx \frac{st}{r}e^{-\delta(t)},
\]
where $0 \le \delta(t) \le 1$.
\end{theorem}

See Appendix~\ref{sec:proof-of-thm-max-es} for the proof.

\section{Consequences of Interventions}
\label{sec:consequences-of-interventions}

I now discuss the implications of Theorem~\ref{thm:max-es} and how an intervention that changes parameters from $(s, r, t)$ to $(s', r', t')$ affects scam operations.

\subsection{Effects on Scam Revenue and Demand for Contact Channels}
\label{sec:effects-on-revenue-and-demand}

If an intervention affects multiple parameters, then the individual effects combine multiplicatively in the total effect on revenue per channel. Thus, even if an intervention's effect on each lever is modest, the effects compound to have significant consequences on scam operations.

I obtain bounds on the effect an intervention that changes the parameters of the scamming environment from $(s, r, t)$ to $(s', r', t')$ has on scammers' revenue and profits. The exact effect depends on scammers' ability to acquire infrastructure in greater quantities.

To obtain an upper bound on the effect an intervention can have, suppose the scammer cannot acquire more contact channels or increase the loss per victim. The ratio of the old revenue under $(s, r, t)$ to the new revenue under $(s', r', t')$ then is
\[
\frac{\mathrm{Revenue}}{\mathrm{Revenue}'} = \frac{\mathrm{E}^\ast(s, r, t)}{\mathrm{E}^\ast(s', r', t')} \approx \frac{s}{s'} \cdot \frac{r'}{r} \cdot \frac{t}{t'} \cdot C,
\]
where $1 \le C \le e$ for $t' \le t$. Suppose $s' = s/2$, $r' = 2r$, and $t' = t/2$. The overall effect is that revenue decreases by a factor of at least $8$.

To obtain a lower bound on the effect, I investigate how scammers would need to scale their infrastructure in order to restore profits after an intervention.

Let $m$ and $m'$ be the number of channels a scammer has access to before and after an intervention, respectively. Suppose that (i) the intervention reduces the expected victims per channel, $\mathrm{E}^\ast(s', r', t') < \mathrm{E}^\ast(s, r, t)$, (ii) the cost of contact channels before and after stays the same or increases, $\kappa' \ge \kappa > 0$, and (iii) this is the only variable cost that changes (see \S\ref{sec:modeling-scam-profits}). Assume also that channels remain profitable after the intervention, \ie, $\ell \cdot \mathrm{E}^\ast(s', r', t') > \kappa'$. If profit is restored then
\[
\overbrace{\ell \cdot \mathrm{E}^\ast(s', r', t') \cdot m'}^{\text{Revenue}} - \overbrace{\kappa'm'}^{\text{Cost}} = \ell \cdot \mathrm{E}^\ast(s, r, t) \cdot m - \kappa m
\]
implies that
\[
\frac{m'}{m} > \frac{\mathrm{E}^\ast(s, r, t)}{\mathrm{E}^\ast(s', r', t')}.
\]

This means that if an intervention reduces revenue per channel by $A\times$, then to restore profits scammers must acquire at least $A\times$ as many channels, increasing infrastructure costs. Because of greater scammer demand for contact channels, the price per channel can also be expected to increase.

Whether increased demand translates into significantly higher costs and a meaningful reduction in scam revenue depends on how the price of channels responds to demand. If supply is highly elastic, the price of contact channels will not significantly increase, scammers will simply acquire more channels, and revenue will likely stay near pre-intervention levels. However, if supply is inelastic, prices will increase significantly, it will not be feasible to acquire all of the necessary channels to maintain profits, and scam revenue will fall by up to $A\times$.

\begin{figure}[t]
\centering
\includegraphics[width=0.55\textwidth]{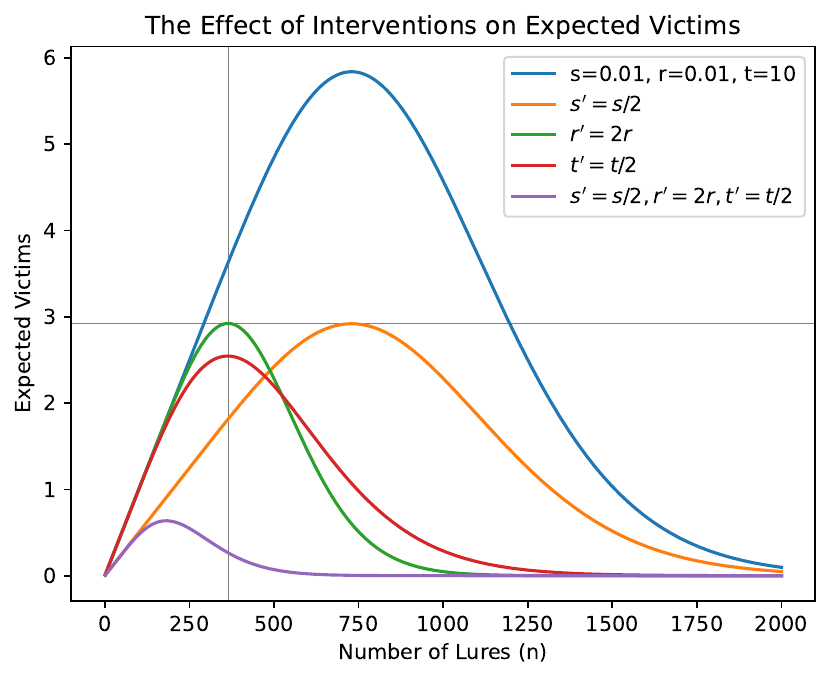}
\caption{The expected victims $\mathrm{E}[S]$ as a function of the number of lures under different parameters. The blue curve represents initial conditions. The yellow ($s' = s/2$), green ($r' = 2r$), and red ($t' = t/2$) curves show the effects of the given intervention. The purple curve shows the effect of all of the interventions together. I plot half the maximum of the blue curve. As predicted by Theorem~\ref{thm:max-es}, this coincides with the maxima of the yellow and green curves, while the maximum of the red curve is slightly less due to the $e^{-\delta(t)}$ factor. I also plot half the optimal number of lures of the blue curve, and as predicted by Theorem~\ref{thm:max-es}, this maximizes the green and red curves. Finally, the multiplicative effect predicted by Theorem~\ref{thm:max-es} appears in the purple curve, where the maximum is $8\times$ lower than that of the blue curve, and the optimal number of lures is one quarter that of the blue curve.}
\label{fig:interventions}
\end{figure}

\subsection{Reducing the Scam Conversion Rate}
\label{sec:reducing-conversion-rate}

Reducing the scam conversion rate affects $\mathrm{E}[S]$ through the amplitude of $\mathrm{E}[S]$, but does not affect the optimal number of lures each channel can send. Say an intervention halves the scam conversion rate, \ie, $s' = s/2$. From Theorem~\ref{thm:max-es} the expected victims, $\mathrm{E}^\ast(s', r, t)$, will approximately halve. To restore profits the scammer must double the number of contact channels they control, and because the optimal number of lures does not change, the total number of lures scammers send must also double.

This agrees with Fig.~\ref{fig:interventions}, where the maximum of the yellow curve is approximately half the maximum of the blue curve, and the shape of the yellow curve versus the blue curve is unchanged.

\subsection{Increasing Reporting}
\label{sec:increasing-reporting}

Increasing the reporting rate affects $\mathrm{E}[S]$ through the amplitude of $\mathrm{E}[S]$ and the optimal number of lures each channel can send. Say an intervention doubles the reporting rate, \ie, $r' = 2r$. From Theorem~\ref{thm:max-es} the expected victims, $\mathrm{E}^\ast(s, r', t)$, and the optimal number of lures will approximately halve. To restore profits the scammer must double the number of contact channels they control, but because the optimal number of lures halves, the total number of lures sent is unchanged.

This agrees with Fig.~\ref{fig:interventions}, where the maximum of the green curve is approximately half the maximum of the blue curve, and the peak of the green curve occurs at half the lures of the blue curve.

\subsubsection{Centralizing Reporting}
\label{sec:centralizing-reporting}

In the real world, users must report to a relevant authority in order for the report to count toward the ban threshold. There may even be multiple relevant authorities. The effect of this is that even if there are $t$ reports in aggregate, each database may have fewer than $t$ reports.

I extend the model (see Appendix~\ref{sec:centralization}) by considering the reporting rate to be the product of the base reporting rate, $r$, and a centralization factor $c$, which captures the probability that a report is made to the relevant authority. This becomes a multiplier on $r$ wherever $r$ appears: the optimal number of lures becomes $n^\ast \approx \lambda^\ast(t)/(cr)$ and the maximum expected victims becomes $\mathrm{E}^\ast(s,c,r,t) \approx \frac{st}{cr}e^{-\delta(t)}$. Interventions that increase centralization affect scams in the same way as increasing the reporting rate.

\subsection{Reducing the Ban Threshold}
\label{sec:reducing-ban-threshold}

Like reporting, reducing the ban threshold affects $\mathrm{E}[S]$ through the amplitude of $\mathrm{E}[S]$ and the optimal number of lures each channel can send. Say an intervention halves the ban threshold, \ie, $t' = t/2$. From Theorem~\ref{thm:max-es} the expected victims, $\mathrm{E}^\ast(s, r, t')$, and the optimal number of lures will approximately halve. To restore profits the scammer must double the number of contact channels they control, but the total number of lures sent is unchanged.

Again, this agrees with Fig.~\ref{fig:interventions}, where the maximum of the red curve is less than half the maximum of the blue curve. Like reporting, the ban threshold affects $\mathrm{E}[S]$ through the optimal number of lures sent, so the shape is similar to the reporting curve.

\subsubsection{False Positives and the Ban Threshold}
\label{sec:false-positives}

The ban threshold encodes a platform's tolerance for banning innocent channels. If a platform's false positive tolerance is low, then the threshold must be high to meet that tolerance. Therefore one way to reduce the ban threshold is to reduce the false positive rate for individual reports.

I express the threshold $t$ as a function of the target false positive rate, $p_{\mathsf{fp}}$, and the expected number of reports an innocent channel receives, $\mu$, which I model with the Poisson distribution.

\begin{lemma}[Reliability-Threshold Relation]
\label{lem:reliability-threshold}
Let $R_{\mathsf{fp}} \sim \mathrm{Poisson}(\mu)$ be a random variable denoting the number of false reports an innocent channel receives, and let $K = \ln(1/p_{\mathsf{fp}})$ with $K > \mu$. Then $\Pr[R_{\mathsf{fp}} \ge t] \leq p_{\mathsf{fp}}$ for all
\[
t \ge \frac{K - \mu}{W_0\left(\frac{K - \mu}{e\mu}\right)},
\]
where $W_0$ is the principal branch of the Lambert $W$ function.
\end{lemma}

See Appendix~\ref{sec:proof-of-lem-reliability-threshold} for the proof.

If an intervention reduces the expected false reports by a factor of $a$, \ie, $\mu' = \mu / a$, then because $W_0(ax) \approx W_0(x) + \ln(a)$ for large arguments,
\[
\frac{m'}{m} \propto \frac{t}{t'} \approx 1 + \frac{\ln(a)}{W_0\left(\frac{K - \mu}{e\mu}\right)}.
\]

Asymptotically, $a$ enters the equation only as a logarithmic addition, so reducing false positives is a weak lever compared to $s$, $r$, and $c$.

As an example, say that the target false positive rate is $p_{\mathsf{fp}} = 0.001$ and the expected number of false reports per channel is $\mu = 1$. Inverting the Poisson CDF and recovering an exact threshold, I obtain $t = 6$ (the bound of Lemma~\ref{lem:reliability-threshold} certifies $t \ge 6.63$). If an intervention halves the expected number of false reports per channel, \ie, $\mu' = 1 / 2$, then $t' = 5$ (the bound certifies $t' \ge 4.96$). The overall effect then is that even though the intervention halved the expected number of false reports, to restore profits the scammer must acquire only $1.2\times$ as many contact channels.

\section{Discussion}
\label{sec:discussion}

In this section, I discuss the implications this work has for designing effective anti-scamming strategies.

The most important result of this paper is that if an intervention affects multiple parameters---the scam conversion rate, reporting rate, centralization of reporting, and false positive rate---the individual effects combine multiplicatively in the total effect on scam revenue per channel, demand for contact channels, and infrastructure costs. Interventions affecting multiple parameters could therefore significantly reduce the profitability of scams.

\subsection{Prioritize Reporting over Education}
\label{sec:prioritize-reporting}

I model the effects of two primary interventions: reducing the probability of being scammed and improving reporting. I believe that improving reporting will be the more effective of the two.

The main way to reduce scam conversion rates is through education. Effective education has proven difficult, with Ho \etal\ \cite{2025-ho-phishing-training} finding that education made little or no difference in reducing phishing click-through rates across a large organization. The reasons are structural. Effective education must first reach those who need it. This is difficult because the cohort of people who cannot recognize scams is small. A general education campaign is therefore likely to waste the majority of resources on people who can already recognize scams. It must then train people who cannot recognize scams to recognize them, and the effect must be persistent as scams evolve. Effective education has already proven difficult and will only become more difficult as AI makes scams more believable.

Reporting is structurally an easier intervention to make. Almost everyone can already recognize scams. The challenge is to convince them to report scams, report them to the relevant authority, and report truthfully.

\subsection{Focus Reporting on Financial Infrastructure}
\label{sec:focus-on-financial-infrastructure}

Reporting interventions are most effective against infrastructure where the supply is inelastic, meaning the price is sensitive to demand. Scams are not run with a single type of channel---phone numbers, social media accounts, and money-laundering accounts are all used in conjunction at different points in the scam. In general, the supply of communication infrastructure is more elastic\footnote{Dek \etal\ \cite{2025-dek-online-manipulation-economy} found SMS verification prices climb ahead of national elections, signaling demand spikes from online influence campaigns. Because verifications depend on a supply of unused phone numbers to receive security codes, this suggests that even the supply of phone numbers---the canonical example of highly elastic infrastructure---is less elastic than assumed.} than the supply of financial infrastructure \cite{2011-levchenko-click-trajectories}.

Reporting elastic channels is unlikely to be effective, because scammers will simply replace banned phone numbers, social media accounts, and emails at similar costs. Even if the costs of communication infrastructure rise significantly, they remain small relative to the other costs of scam operations. For example, fresh, phone-verified Gmail accounts only cost about \$1 each \cite{2026-bulk-gmail-account-vendor}, while phone verifications of online accounts using US SIM cards cost only \$0.26 \cite{2025-dek-online-manipulation-economy}.

Reporting financial infrastructure, however, is likely to be very effective. Scammers' financial accounts can be worth hundreds of dollars on the black market \cite{2026-group-ib-cloud-phones, 2026-group-ib-mule-account-creation, 2025-sumsub-money-mules} and are difficult to acquire. Spokoyny \etal\ \cite{2025-spokoyny-victim-as-a-service} found that scammers reveal financial accounts and crypto wallets about 70 messages into a conversation. An intervention that increases reporting of these accounts could significantly increase the required number and price of these accounts, causing scam costs to increase dramatically.

Money-laundering accounts have a further advantage as a target: anti-money-laundering law already obligates financial institutions to identify and close accounts used to move illicit proceeds. Reporting interventions aimed at this infrastructure reinforce an existing legal mandate rather than relying on the voluntary enforcement that governs communication channels.

\subsection{Increase Centralization of Reporting}
\label{sec:increase-centralization}

My modeling suggests centralization is a high-leverage intervention, so building a single system to ingest fraud intelligence across all types of scam infrastructure is a high priority. Increased centralization is likely to also increase reporting. In GASA's 2026 State of Scams in the US \cite{2026-gasa-us-state-of-scam}, 36\% of respondents said they did not report a scam simply because they were unsure where to report it. A central, well-publicized repository for scam intelligence would eliminate this issue.

Current scam reporting is highly decentralized, so signal is spread across many non-communicating systems. Each social media platform has its own reporting system. Even among law enforcement agencies, there are multiple different databases for reporting fraud. And while financial institutions have better data-sharing practices, they only see transactions, at which point money is often unrecoverable and the financial accounts involved have rotated.

\subsection{A Bound on Scam Profitability}
\label{sec:bound-on-profitability}

My modeling suggests reporting could be highly effective, especially when applied to high-cost, inelastic infrastructure. But could it actually make scams unprofitable?

Using the model I developed, I obtain an upper bound on the reporting rate that will render YouTube crypto scams unprofitable. Where possible I use exact statistics, but when data does not exist I overestimate parameters in a way that favors scammers. The bounds I obtain therefore overestimate the real rates required to make scams unprofitable.

The average loss per victim is \$20,600 \cite{2025-fbi-ic3-internet-crime-report} and the rate at which viewing a scam livestream converts into a victim is 0.0039\% \cite{2024-liu-give-and-take}. The cost of a hijacked YouTube account is between \$3 and \$4,000 \cite{shen2021youtubephishing} with a median cost of \$759 \cite{2024-beluri-sale-of-social-accounts}. Other parameters are not publicly available, so I make assumptions that favor scammers. (i) With AI, scam labor costs will go down, so I assume scammers' only costs are infrastructure. In reality scammers incur many other costs. (ii) Prices for scam infrastructure will increase with demand. I assume prices will remain the same. (iii) The average number of false reports per innocent account is near zero. I assume each innocent account gets one false report on average, drastically underestimating the reliability of reports. (iv) I assume the platform tolerates a false positive rate of only one in a thousand.

Assuming reports are acted upon in a timely manner, with the above parameters, scams will become unprofitable if 85\% of viewers report a \$3 account, while with a \$4,000 account just 0.06\% must report. With the median account costing \$759, scams would no longer be profitable if just 0.34\% of viewers reported. 

This analysis suggests that reporting expensive, inelastic infrastructure is a realistic pathway to making scams unprofitable. Because I underestimate scammers' costs, these figures represent an upper bound on the reporting rates required to make scams unprofitable.

\section{Conclusion}
\label{sec:conclusion}

In this paper I have developed a simple model of scam profits under worst-case assumptions about how AI will affect scam costs.

Using this model, I showed that increasing reporting and centralization, and reducing false reports, have a multiplicative effect on revenue per scam contact channel, costs of infrastructure to maintain profits, and scammers' demand for infrastructure. This multiplicative effect gives us an advantage over scammers, even if AI increases scam conversion rates in the coming years.

The challenge is to design a reporting system that improves all three parameters simultaneously and quickly distills reports into actionable intelligence. It is easy to design a reporting system that improves one or two parameters at the expense of the others. For example, a system that made reporting scams extremely easy would attract more reports, increasing both true reporting and centralization. But it would also attract more false reports, diminishing the efficacy of the system.

If we could build such a system, my analysis suggests it could offer a realistic path to turning the tide against scammers. And if the effect were strong enough, it is conceivable that it could even make scams unprofitable.

\section*{Acknowledgements}
\label{sec:acknowledgements}

Thank you to Dr. Rajvardhan Oak and Camille Fredrickson. Their feedback and suggestions greatly improved the presentation and content of this paper.

\bibliographystyle{IEEEtran}
\bibliography{refs}

@inproceedings{2025-oak-anna,
  author       = {Oak, Rajvardhan and Shafiq, Zubair},
  title        = {{`Hello, is this Anna?': Unpacking the Lifecycle of Pig-Butchering Scams}},
  booktitle    = {USENIX SOUPS},
  year         = {2025}
}

@article{2025-dek-online-manipulation-economy,
  author       = {Dek, A. and Kyrychenko, Y. and van der Linden, S. and Roozenbeek, J.},
  title        = {{Mapping the online manipulation economy}},
  journal      = {Science},
  year         = {2025}
}

@misc{cfa2026_scam_economy,
  author      = {{Consumer Federation of America}},
  title       = {{The Scam Economy: The True Cost of Online Scams and Crimes in America}},
  year        = {2026}
}

@misc{2025-fbi-ic3-internet-crime-report,
    author = {{FBI IC3}},
    title = {{Internet Crime Report 2025}},
    year = {2026}
}

@misc{2025-gasa-global-state-of-scams,
    author = {{Global Anti-Scam Alliance (GASA)}},
    title = {{Global State of Scams 2025}},
    year = {2026}
}

@misc{2026-chainalysis-fraud-report,
    author = {{Chainalysis}},
    title = {{Record \$17 Billion Estimated Stolen in Crypto Scams and Fraud in 2025}},
    year = {2026}
}

@inproceedings{2024-liu-give-and-take,
    author = {Liu, Enze and others},
    title = {{Give and Take: An End-to-End Investigation of Giveaway Scam Conversion Rates}},
    booktitle = {{ACM IMC}},
    year = {2024}
}

@inproceedings{2025-ho-phishing-training,
  title={{Understanding the efficacy of phishing training in practice}},
  author={Ho, Grant and Mirian, Ariana and Luo, Elisa and Tong, Khang and Lee, Euyhyun and Liu, Lin and Longhurst, Christopher A. and Dameff, Christian and Savage, Stefan and Voelker, Geoffrey M.},
  booktitle={2025 IEEE Symposium on Security and Privacy (SP)},
  year={2025}
}

@misc{2026-group-ib-cloud-phones,
  author       = {Anarkulov, Anvar and Turakhujaev, Khumoyun},
  title        = {{Cloud Phones Explained: Why They Are Becoming a Major Fraud Threat for Banks}},
  howpublished = {Group-IB Blog},
  year         = {2026},
  month        = {mar},
  url          = {https://www.group-ib.com/blog/cloud-phones-invisible-threat/},
  note         = {Accessed 2026-07-08}
}

@misc{2026-group-ib-mule-account-creation,
  author       = {Grabko, Alexander and Angelopoulos, Konstantinos},
  title        = {{Anatomy of a Fraud Operation: Mule Account Creation on B2B Fintech Platforms in France}},
  howpublished = {Group-IB Blog},
  year         = {2026},
  url          = {https://www.group-ib.com/blog/french-fintech-mule-accounts/},
  note         = {Accessed 2026-07-08}
}

@misc{2025-sumsub-money-mules,
  author       = {Abramova, Alisa and Prior, Ed},
  title        = {What Is a Money Mule? Red Flags, Examples, and Prevention in 2025},
  howpublished = {Sumsub Blog},
  year         = {2025},
  url          = {https://sumsub.com/blog/money-muling/},
  note         = {Accessed 2026-07-08}
}

@misc{2025-lexisnexis-true-cost-of-fraud-north-america,
  author      = {{LexisNexis Risk Solutions}},
  title       = {{True Cost of Fraud Study 2025 North America}},
  year        = {2025}
}

@misc{shen2021youtubephishing,
  author       = {Shen, Ashley},
  title        = {{Phishing Campaign Targets YouTube Creators with Cookie Theft Malware}},
  howpublished = {Google Threat Analysis Group Blog},
  year         = {2021},
}

@inproceedings{2026-gressel-ai-romance-scams,
  author    = {Gressel, Gilad and others},
  title     = {{Love, Lies, and Language Models: Investigating AI's Role in Romance-Baiting Scams}},
  booktitle = {USENIX Security},
  year      = {2026}
}

@inproceedings{2008-kanich-spamalytics,
  author    = {Kanich, Chris and Kreibich, Christian and Levchenko, Kirill and Enright, Brandon and Voelker, Geoffrey M. and Paxson, Vern and Savage, Stefan},
  title     = {{Spamalytics: An Empirical Analysis of Spam Marketing Conversion}},
  booktitle = {ACM CCS},
  year      = {2008}
}

@inproceedings{2011-levchenko-click-trajectories,
  author    = {Levchenko, Kirill and others},
  title     = {{Click Trajectories: End-to-End Analysis of the Spam Value Chain}},
  booktitle = {IEEE Symposium on Security and Privacy},
  year      = {2011}
}

@misc{2024-beluri-sale-of-social-accounts,
  author       = {Beluri, Mario and others},
  title        = {{Exploration of the Dynamics of Buy and Sale of Social Media Accounts}},
  howpublished = {arXiv:2412.14985},
  year         = {2024}
}

@misc{2025-spokoyny-victim-as-a-service,
  author       = {Spokoyny, Daniel and others},
  title        = {{Victim as a Service: Designing a System for Engaging with Interactive Scammers}},
  howpublished = {arXiv:2510.23927},
  year         = {2025}
}

@article{2026-heiding-ai-voice-phishing,
  author  = {Heiding, Fred and others},
  title   = {{Evaluating AI Models' Capability to Automate Voice Phishing Attacks}},
  journal = {Expert Systems with Applications},
  year    = {2026}
}

@misc{2026-gasa-us-state-of-scam,
  author      = {{Global Anti-Scam Alliance (GASA)}},
  title       = {{State of Scams in the United States 2026}},
  year        = {2026}
}

@misc{2026-bulk-gmail-account-vendor,
  author       = {{Bulk Accounts Buy}},
  title        = {{Bulk Gmail Accounts for Sale}},
  howpublished = {Vendor listing},
  note         = {Accessed 2026-08-30}
}

@misc{2026-ftc-jury-duty-scam,
  author       = {{Federal Trade Commission}},
  title        = {{Ignore calls, texts, and emails threatening to arrest you for missing jury duty}},
  howpublished = {FTC Consumer Alerts},
  year         = {2026},
  month        = jun,
  url          = {https://consumer.ftc.gov/consumer-alerts/2026/06/ignore-calls-texts-and-emails-threatening-arrest-you-missing-jury-duty},
  note         = {Accessed 2026-09-01}
}

\clearpage

\appendix
\section*{Appendix}

\section{Proof of Theorem~\texorpdfstring{\ref{thm:max-es}}{1}}
\label{sec:proof-of-thm-max-es}

I make the following definitions for convenience:
\begin{itemize}
  \item $P(\lambda, t) = \Pr[\mathrm{Poisson}(\lambda) \leq t-1]$,
  \item $p_j(\lambda) = e^{-\lambda}\frac{\lambda^j}{j!}$,
  \item $f_t(\lambda) = \lambda P(\lambda, t)$,
  \item $g(t) = \max \limits_{\lambda > 0} f_t(\lambda)$, and $\lambda^\ast(t) = \arg\max\limits_{\lambda>0}f_t(\lambda)$, so $g(t) = f_t(\lambda^\ast(t))$.
\end{itemize}

\begin{rthm}
Let $\mathrm{E}^\ast(s, r, t)$ be the maximum value of $\mathrm{E}[S \mid n]$. Then the maximum of $\mathrm{E}[S \mid n]$ is attained at $n^\ast \approx \frac{\lambda^\ast(t)}{r}$, where $\lambda^\ast(t)$ maximizes $\lambda \Pr[\mathrm{Poisson}(\lambda) < t]$, and
\[
\mathrm{E}^\ast(s, r, t) \approx \frac{st}{r}e^{-\delta(t)},
\]
where $0 \le \delta(t) \le 1$.
\end{rthm}

Let $\mathrm{E}^\ast(s,r,t) = \max\limits_{n > 0} sn \Pr[R \leq t - 1]$. When $n$ is large and $r$ is small, $\mathrm{Binomial}(n, r) \approx \mathrm{Poisson}(rn)$. Let $\lambda = rn$. Substituting this into the above, we solve
\[
\mathrm{E}^\ast(s,r,t) \approx \frac s r \max\limits_{\lambda > 0} \lambda \Pr[\mathrm{Poisson}(\lambda) \leq t - 1].
\]

By Lemma~\ref{lem:threshold-scaling} we know that for $t' < t$, $g(t) = g(t') \frac{t}{t'} \prod\limits_{u = t'}^{t-1} A(u)$. We know that $g(1) = \max\limits_{\lambda > 0} \lambda e^{-\lambda} = \frac{1}{e}$, so for $t > 1$, $g(t) = \frac{t}{e} \prod\limits_{u = 1}^{t-1} A(u)$.

We now show that $\prod\limits_{u = 1}^{\infty} A(u) = e$. Consider $f_t(\lambda) = \sum\limits_{j = 0}^{t - 1} \lambda p_j(\lambda)$. Observe that $\lambda p_j(\lambda) = (j+1)p_{j+1}(\lambda)$, so reindexing with $k = j+1$ gives $f_t(\lambda) = \sum\limits_{k = 1}^t k p_k(\lambda)$. Therefore, for all $\lambda$
\begin{align*}
    f_t(\lambda) = & \sum\limits_{k = 1}^t k p_k(\lambda) \\
    \le & t \sum\limits_{k = 1}^t p_k(\lambda) \\
    < & t.
\end{align*}

Because $\lambda^\ast(t)$ is a particular value of $\lambda$, we have that $g(t) < t$. So
\begin{align*}
    g(t) = \frac{t}{e} \prod\limits_{u = 1}^{t-1} A(u) < & t \\
    \prod\limits_{u = 1}^{t-1} A(u) < & e
\end{align*}

We now show that $g(t)/t \to 1$ as $t$ increases. We know that $g(t) < t$, so $g(t)/t < 1$. From the other side, we know that $g(t) \ge \lambda P(\lambda, t)$ for all $\lambda$, so let $\lambda = t - t^{2/3}$. We use Chebyshev's inequality to provide a lower bound on $P(\lambda, t)$. Let $X \sim \mathrm{Poisson}(\lambda)$, so $\Pr[X \ge t] = 1 - P(\lambda, t)$. Because $t - \lambda = t^{2/3}$, the event $X \ge t$ implies $|X - \lambda| \ge t^{2/3}$, so $\Pr[X \ge t] \leq \Pr[|X - \lambda| \ge t^{2/3}]$. By Chebyshev's inequality,
\[
\Pr[|X - \lambda| \ge t^{2/3}] \le \frac{\mathrm{Var}[X]}{t^{4/3}} = \frac{\lambda}{t^{4/3}} \le t^{-1/3},
\]
so $P(\lambda, t) \ge 1 - t^{-1/3}$. Substituting this, we have
\[
1 - 2t^{-1/3} \le \frac{g(t)}{t} < 1,
\]
which implies that as $t \to \infty$, $g(t)/t \to 1$. Finally, $\prod\limits_{u = 1}^{t-1} A(u) = e \frac{g(t)}{t},$ so because $g(t)/t \to 1$
\[
\prod\limits_{u = 1}^{\infty} A(u) = e.
\]

We let $\delta(t) = \sum\limits_{u = t}^\infty \ln A(u)$. Because $A(u) \ge 1$ (by Lemma~\ref{lem:threshold-scaling}), $\delta(t) \ge 0$, and because $\prod\limits_{u = 1}^{\infty} A(u) = e$, $\delta(t) \leq 1$. After algebra we obtain
\[
g(t) = t e^{-\delta(t)}.
\]

Substituting this back into our equation for $\mathrm{E}^\ast(s, r, t)$ we have
\[
\mathrm{E}^\ast(s, r, t) \approx \frac{st}{r}e^{-\delta(t)}.
\]

\subsection{Proof of Lemma~\texorpdfstring{\ref{lem:threshold-scaling}}{2}}
\label{sec:proof-of-lem-threshold-scaling}

\begin{lemma}[Threshold Scaling of the Poisson Maximum]
\label{lem:threshold-scaling}
Let $t' < t$ and $g(t) = \max \limits_{\lambda > 0} \lambda \Pr[\mathrm{Poisson}(\lambda) \leq t-1]$. Then
\[
\frac{g(t)}{g(t')} = \frac{t}{t'} \prod\limits_{u = t'}^{t-1} A(u),
\]
where $A(t) = \frac{g(t+1)}{g(t) \left(1 + \frac{1}{t}\right)} \ge 1$.
\end{lemma}

We begin by deriving a relation for $g(t+1)/g(t)$. Observe that by definition $g(t+1) \ge f_{t+1}(\lambda^\ast(t))$. Because $P$ is a CDF, $P(\lambda, t + 1) = P(\lambda, t) + p_t(\lambda)$, so $f_{t+1}(\lambda^\ast(t)) = \lambda^\ast(t) P(\lambda^\ast(t), t+1) = \lambda^\ast(t)[P(\lambda^\ast(t), t) + p_t(\lambda^\ast(t))]$. Factoring out $P(\lambda^\ast(t), t)$, we have
\begin{align*}
    g(t+1) \ge f_{t+1}(\lambda^\ast(t)) = & \lambda^\ast(t)P(\lambda^\ast(t), t) \left[1 + \frac{p_t(\lambda^\ast(t))}{P(\lambda^\ast(t), t)}\right] \\
    = & g(t) \left[1 + \frac{p_t(\lambda^\ast(t))}{P(\lambda^\ast(t), t)}\right].
\end{align*}

We now show that $\frac{p_t(\lambda^\ast(t))}{P(\lambda^\ast(t), t)} = \frac 1 t$. First, observe that $\frac{p_t(\lambda)}{p_{t - 1}(\lambda)} = \frac{e^{-\lambda}\lambda^{t}/t!}{e^{-\lambda}\lambda^{t-1}/(t-1)!} = \frac{\lambda}{t}$. So
\begin{equation}
\label{eq:pt-recurrence}
p_t(\lambda) = \frac{\lambda}{t} p_{t-1}(\lambda)
\end{equation}

Now $\lambda^\ast(t)$ is the maximizer of $f_t$, so $f'_t(\lambda^\ast(t)) = 0$. Differentiating $f_t(\lambda)$ with respect to $\lambda$ we get $f'_t(\lambda) = P(\lambda, t) - \lambda p_{t-1}(\lambda)$, and therefore
\begin{equation}
\label{eq:foc}
P(\lambda^\ast(t), t) = \lambda^\ast(t) p_{t-1}(\lambda^\ast(t))
\end{equation}

Substituting Eq.~\ref{eq:pt-recurrence} and Eq.~\ref{eq:foc}, we have that $\frac{p_t(\lambda^\ast(t))}{P(\lambda^\ast(t), t)} = \frac 1 t$. Therefore,
\[
g(t+1) \ge g(t)\left(1 + \frac{1}{t}\right).
\]

We define $A(t)$ such that $g(t+1) = g(t)\left(1 + \frac{1}{t}\right)A(t)$. So
\[
\frac{g(t+1)}{g(t)} = A(t)\left(1 + \frac{1}{t}\right).
\]

Because $A(t) = \frac{g(t+1)}{g(t) \left(1 + \frac{1}{t}\right)}$ and $\frac{g(t+1)}{g(t) \left(1 + \frac{1}{t}\right)} \ge 1$, $A(t) \ge 1$.

We now prove the result. Observe that for $t' < t$
\begin{align*}
    \frac{g(t)}{g(t')} = &\frac{g(t' + 1)}{g(t')} \cdot \frac{g(t' + 2)}{g(t' + 1)} \cdot \cdots \cdot \frac{g(t)}{g(t - 1)} \\
    = & \frac{t' + 1}{t'}A(t') \cdot \frac{t' + 2}{t' + 1}A(t' + 1) \cdot \cdots \cdot \frac{t}{t - 1}A(t-1) \\
    = & \frac{t}{t'} \prod\limits_{u = t'}^{t - 1}A(u).
\end{align*}

\section{Centralization}
\label{sec:centralization}

We extend the model by requiring that a report is not only filed, but filed with the single authority able to ban the channel, \eg, the account is managed by Instagram, which alone can ban it. Let $c = \Pr[\mathrm{db} \mid \mathrm{report}]$ denote the probability that a filed report reaches this authority. The probability that a lure produces a report in the relevant database is then
\[
\Pr[\mathrm{report} \land \mathrm{db}] = \Pr[\mathrm{db} \mid \mathrm{report}]\,\Pr[\mathrm{report}] = cr,
\]
which substitutes for $r$ throughout Theorem~\ref{thm:max-es}. The optimal lure count becomes $n^\ast \approx \frac{\lambda^\ast(t)}{cr}$, the maximum expected victims becomes
\[
\mathrm{E}^\ast(s, c, r, t) \approx \frac{st}{cr}e^{-\delta(t)},
\]
and
\[
\frac{m'}{m} > \frac{s}{s'} \cdot \frac{c'}{c} \cdot \frac{r'}{r} \cdot \frac{t}{t'} \cdot e^{\delta(t') - \delta(t)},
\]
under the same assumptions as \S\ref{sec:effects-on-revenue-and-demand}.

\section{Proof of Lemma~\texorpdfstring{\ref{lem:reliability-threshold}}{1}}
\label{sec:proof-of-lem-reliability-threshold}

\begin{rlemA}
Let $R_{\mathsf{fp}} \sim \mathrm{Poisson}(\mu)$ be a random variable denoting the number of false reports an innocent channel receives, and let $K = \ln(1/p_{\mathsf{fp}})$ with $K > \mu$. Then $\Pr[R_{\mathsf{fp}} \ge t] \leq p_{\mathsf{fp}}$ for all
\[
t \ge \frac{K - \mu}{W_0\left(\frac{K - \mu}{e\mu}\right)},
\]
where $W_0$ is the principal branch of the Lambert $W$ function.
\end{rlemA}

Let $\theta > 0$. Because $x \mapsto e^{\theta x}$ is increasing, $\{R_{\mathsf{fp}} \ge t\} = \{e^{\theta R_{\mathsf{fp}}} \ge e^{\theta t}\}$. By Markov's inequality and the Poisson moment generating function $\mathrm{E}[e^{\theta R_{\mathsf{fp}}}] = e^{\mu(e^{\theta} - 1)}$,
\[
\Pr[R_{\mathsf{fp}} \ge t] \leq e^{-\theta t}\, \mathrm{E}[e^{\theta R_{\mathsf{fp}}}] = \exp\left(\mu(e^{\theta} - 1) - \theta t\right).
\]

The bound holds for every $\theta > 0$, so we minimize the exponent over $\theta$. The derivative is
\[
\frac{d}{d\theta} \left[ \mu(e^{\theta} - 1) - \theta t \right ] = \mu e^{\theta} - t,
\]
so when it is zero, $\theta = \ln(t/\mu)$, which is positive when $t > \mu$. Substituting this back,
\[
\Pr[R_{\mathsf{fp}} \ge t] \leq \exp\left(t - \mu - t\ln(t/\mu)\right) = e^{-\mu}\left(\frac{e\mu}{t}\right)^{t}.
\]

Therefore we solve for $t$ such that $e^{-\mu}(e\mu/t)^t \leq p_{\mathsf{fp}}$. Taking the natural log of both sides and rearranging, we have
\[
t \ln\left(\frac{t}{e\mu}\right) \ge K - \mu.
\]

Let $h(t) = t\ln(t/(e\mu))$. Because $K > \mu$, we need $h(t) > 0$. This is the region $t > e\mu$, and on this interval $h$ is positive and increasing, so all $t \ge t^\ast$ satisfy $h(t) \ge K - \mu$, where $h(t^\ast) = K - \mu$. To solve for $t^\ast$, substitute $v = \ln(t/(e\mu))$, so $t = e\mu e^{v}$ and
\begin{align*}
h(t^\ast) = e\mu\, v e^{v} = & K - \mu \\
v e^{v} = & \frac{K - \mu}{e \mu}
\end{align*}

The right side is positive, so $v = W_0\left(\frac{K - \mu}{e\mu}\right)$, and
\[
t^\ast = e\mu e^{v} = \frac{K - \mu}{W_0\left(\frac{K - \mu}{e\mu}\right)}.
\]

Therefore for all $t \ge \frac{K - \mu}{W_0\left(\frac{K - \mu}{e\mu}\right)}$, $\Pr[R_{\mathsf{fp}} \ge t] \leq p_{\mathsf{fp}}$.

\end{document}